\documentclass[useAMS,usenatbib]{mn2e}

\usepackage{amsmath,array,amsfonts}
\usepackage{amssymb}
\usepackage{graphicx}
\usepackage{natbib}
\usepackage{pstricks}
\usepackage{verbatim}
\usepackage{multirow}

\newcommand{\ltsima} {$\; \buildrel < \over \sim \;$}
\newcommand{\gtsima} {$\; \buildrel > \over \sim \;$}
\newcommand{\lta} {\lower.5ex\hbox{\ltsima}}
\newcommand{\gta} {\lower.5ex\hbox{\gtsima}}

\title[Multi-resolution internal template cleaning]{Multi-resolution internal template cleaning: An application to the \textit{Wilkinson Microwave Anisotropy Probe} 7-yr polarization data}

\author[R. Fern\'{a}ndez-Cobos et al.]{R. Fern\'{a}ndez-Cobos$^1$$^,$$^2$\thanks{e-mail:
cobos@ifca.unican.es}, P. Vielva$^1$, R.B. Barreiro$^1$, E. Mart\'{i}nez-Gonz\'{a}lez$^{1}$\\
$^1$     Instituto de F\'isica de Cantabria, CSIC-Universidad de Cantabria, Avda. de los Castros s/n, 39005 Santander, Spain.\\
$^2$     Dpto. de F\'isica Moderna, Universidad de Cantabria, Avda. los Castros s/n, 39005 Santander, Spain.}

\date{Accepted  Received ; in original form }

\begin{document}

\maketitle

\begin{abstract}
Cosmic microwave background (CMB) radiation data obtained by different
experiments contain, besides the desired signal, a superposition of
microwave sky contributions. We present a fast and robust method,
using a wavelet decomposition on the sphere, to recover the CMB signal
from microwave maps. An application to \textit{WMAP} polarization data
is presented, showing its good performance particularly in very
polluted regions of the sky. The applied wavelet has the advantages of
requiring little computational time in its calculations, being adapted
to the \textit{HEALPix} pixelization scheme, and offering the
possibility of multi-resolution analysis. The decomposition is
implemented as part of a fully internal template fitting method,
minimizing the variance of the resulting map at each
scale. Using a $\chi^2$ characterization of the noise, we find
  that the residuals of the cleaned maps are compatible with those
  expected from the instrumental noise. The maps are also comparable
  to those obtained from the \textit{WMAP} team, but in our case
  we do not make use of external data sets. In addition, at low
  resolution, our cleaned maps present a lower level of noise. The E-mode power
spectrum $C_{\ell}^{EE}$ is computed at high and low resolution; and a
cross power spectrum $C_{\ell}^{TE}$ is also calculated from the
foreground reduced maps of temperature given by \textit{WMAP} and our
cleaned maps of polarization at high resolution. These spectra are
consistent with the power spectra supplied by the \textit{WMAP}
team. We detect the E-mode acoustic peak at $\ell \sim 400$, as
predicted by the standard $\Lambda CDM$ model. The B-mode power
spectrum $C_{\ell}^{BB}$ is compatible with zero.
\end{abstract}

\section{Introduction}
Component separation is a critical aspect in the analysis of cosmic microwave background (CMB) data. A good characterization of the data is a prerequisite to the adequate estimation of cosmological parameters. This need becomes crucial when, as happens in B-mode detection experiments, foreground amplitudes are well above the signal \citep[e.g.,][]{Tucci2005}. Two physical galactic processes are the major contaminants to CMB polarized signal: synchrotron radiation and thermal dust. Both appear at large scales, are highly anisotropic and the spatial variation of their emissivity is smooth. Besides, extragalactic emission also contaminates this cosmological signal: point sources and clusters are compact objects, roughly isotropically distributed in the sky and every single object has a particular frequency dependence. Most of the component separation methods take into account only diffuse components, assuming that we are previously masking the brightest point sources or subtracting them by, typically, fitting approaches \citep[see][for a recent review]{Herranz2010}.

Current and future experiments \citep{QUIJOTE2008, QUaD2009, CBI2009, POLARBEAR2010, PAPPA2006, Grainger2008, BRAIN2008} are able to measure CMB polarization anisotropies with such precision that foreground contamination have become the major limitation when we try to analyze the data. This is the principal reason to invest effort and time in developing new techniques for separating components. The goal of all the proposed methods is to separate or, at least, to identify CMB anisotropies from the other components. The range of proposals includes internal linear combinations (ILC), Bayesian methods and independent component analysis \citep[see][for a recent review]{Del2007}. 

There is abundant literature that includes applications of the various methods related to some polarization experiments in vogue as, for instance, \textit{PLANCK} \citep{Leach2008, Efsta2009, Bet2009, Baccigalupi2004} and \textit{WMAP} \citep{Gold2011, Del2009, Kim2009, Bonaldi2007, Maino2007, Eriksen2006, Eriksen2008}.

The method that we present in this paper is situated in the context of the internal linear combinations and it is a bet for a template cleaning in which coefficients are fitted in the space of a particular wavelet that enables a multi-resolution analysis. A fitting by scales allows, in practice, some effective variation of the coefficients in the sky, which is an advantage over the template cleaning in real space. In this sense, our approach based on wavelet space effectively lies in between standard linear combination techniques applied in the real space and more sophisticated parametric methods \citep[e.g.,][]{Eriksen2006, Eriksen2008, Stivoli2010}.

Our approach is a fast procedure that especially shows its effectiveness in polluted regions, such as those that appear in polarization experiments. 

This paper is structured as follows. The methodology is described in detail in Section \ref{sec:Method}. We set out an analysis of the low-resolution polarization \textit{WMAP} data in Section \ref{sec:lowresWMAP}. In section \ref{sec:highresWMAP}, we show the treatment for high-resolution \textit{WMAP} data in order to obtain the $C_{\ell}^{EE}$ and $C_{\ell}^{TE}$ spectra. Finally, we present the conclusions and discussion in section \ref{sec:conclusions}.

\section{Methodology}
\label{sec:Method}
In this work, we present a multi-resolution internal template cleaning (MITC) method for foreground removal. This is the initial step of the map cleaning process in the \textit{SEVEM} method \citep{Martinez2003, Leach2008} to the case of polarization. 

For many purposes, it is a key point to have CMB maps at several frequencies instead of a single map. For instance, it would serve as a consistency check to verify whether any detected feature of the data is actually monochromatic or not (as, for instance, the case for non-gaussianity analysis). 

Another advantage of the method is that we do not need a thorough knowledge of foregrounds, because we obtain all the information to construct different templates from the data. Furthermore, this procedure preserves the original resolution of the CMB component. But the downside is that the internal templates are noisy, so we increase the total noise level when we remove them from the data. This circumstance results, for instance, in an increase in the error bars of the power spectrum at high multipoles. An alternative would be to incorporate external templates, created from data from  other independent observations or based on theoretical arguments. However, the current knowledge of foreground emissions, in polarization, is not substantiated with suitable ancillary data set, and for that reason, this option is not considered in this case. This situation may change in the future with the information expected to be provided by experiments like \textit{PLANCK} \citep{PLANCK2010}, \textit{C-BASS} \citep{CBASS2011} or \textit{QUIJOTE} \citep{QUIJOTE2008}. 

\subsection{The \textit{HEALPix} wavelet}
Wavelets are a powerful tool in signal analysis and are extensively used in many astrophysics applications. Several examples of implementation of component separation methods which employ very diverse wavelets can be found in the literature \citep[e.g.,][]{Ghosh2011, Del2009, Gonz2006, Vielva2003, Hansen2006}. They are localized wave functions, that allow for a multi-resolution treatment of the data. This fact represents an advantage over other component separation methods because it allows us to vary the effective emissivity of foregrounds.

\begin{figure}
\centering
\resizebox{9cm}{!}{\includegraphics{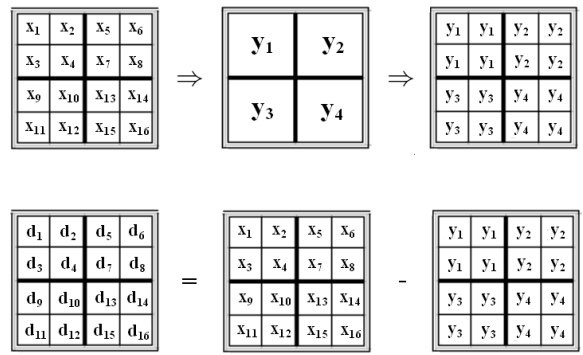}} 
\caption{Outline of construction of the detail coefficients at resolution $j$ ($d_i$) as the substraction of the approximation coefficients at resolution $j$-$1$ ($y_i$), from the approximation coefficients at resolution $j$ ($x_i$).} 
\label{fig:wdetails}
\end{figure}

We use the so-called \textit{HEALPix} wavelet, HW, \citep{Cas2010}, a discrete and orthogonal wavelet that provides a multi-scale decomposition on the sphere adapted to the \textit{HEALPix} pixelization \citep{Gorski2005}. The resolution of a map in the \textit{HEALPix} tessellation is given in terms of the $N_{side}$ parameter, defined so that the number of pixels needed to cover the sphere is $N=12N_{side}^2$. The resolution $j$ of a map is a number such that $2^j=N_{side}$. A CMB map is decomposed in the wavelet coefficient space in a series of maps from the resolution of the original map to the lowest resolution considered. All of these maps, except the lowest resolution one, are called details. The last one is called the approximation, and is constructed by degrading the original map to the appropiate resolution, i.e., to calculate the approximation coefficient at resolution $j$-$1$ at a given position $i$ we take the average of the four daughter pixels at resolution $j$. The way that different detail maps are built is illustrated in figure \ref{fig:wdetails}. At each resolution $j$, detail coefficients are calculated as the substraction of the approximation coefficients at resolution $j$-$1$ from the approximation coefficients at resolution $j$. Both this process and the mathematical formalism of this wavelet is carefully explained in \citet{Cas2010}, where the HW is used to put constraints on the $f_{NL}$ parameter from \textit{WMAP} data. 

In that paper, it is said that the reconstruction of a map $M(x_i)$ can be written as \begin{equation}
\begin{array}{lcl}
M(x_{i}) &=&  {\displaystyle\sum_{p=0}^{n_{j_{0}}-1} \lambda_{j_{0},p} \phi_{j_{0}, p} (x_{i})} \  \  \ + \\
& +&  {\displaystyle\sum_{m = 1}^{4} \sum_{j=j_{0}}^{J-1}\sum_{p=0}^{n_{j}-1} \gamma_{m,j,p} \psi_{m,j,p}(x_{i}) } , 
\end{array}
\end{equation} \\
where $\lambda_{j_{0},\ell}$ and $\gamma_{m,j,\ell}$ are the approximation and detail coefficients respectively, $\phi_{j, p} (x_{i})$ is the scaling function and $\psi_{m,j,p}(x_{i})$ refers to the wavelet functions. The $j$ index takes values from the highest resolution $J$ to the approximation resolution $j_0$. 

The advantage of this wavelet with respect to others, in addition to its straightforward implementation, lies in the speed of the involved operations. The computational time for the wavelet decomposition is of the order of the number of pixels ($\sim N_{pix}$) whereas, for example, for the continous wavelet transform of the spherical Mexican hat wavelet \citep{Martinez2002} or the needlets \citep{Baldi2009} this time is of the order of $\sim N_{pix}^{3/2}$.

\begin{table*}
\centering
\begin{tabular}{cccccc}
\hline
\hline
Frequency band & & Q1 & Q2 & V1 & V2 \\
\hline
\multirow{2}{*}{Detail ($j=4$)} & \textit{Q} Stokes & 0.092 & 0.103 & 0.036 & 0.023 \\ 
 & \textit{U} Stokes & 0.074 & 0.117 & 0.020 & 0.048 \\
\hline
\multirow{2}{*}{Approximation ($j=3$)} & \textit{Q} Stokes & 0.244 & 0.259 & 0.081 & 0.125 \\ 
 & \textit{U} Stokes & 0.241 & 0.236 & 0.085 & 0.112 \\
\hline
\end{tabular}
\caption{Template cleaning coefficients for \textit{Q} and \textit{U} Stokes parameters and DAs for the low resolution case.} \label{tab:DAcoeff} 
\end{table*}

\subsection{Template fitting}
The signal $\widehat{T}_{j}(p)$ at resolution $j$ is constructed by subtracting a linear combination of different templates $t_{ij}$ from the original signal, $T_j$, as follows \begin{equation}
\widehat{T}_{j}(p) = T_{j}(p) - \sum_{i=1}^{N_{t}} \beta_{ij}t_{ij}(p), 
\end{equation} where $N_t$ is the total number of templates and $p$ is a pixel index.

An internal template is formed as the difference of two maps of the same resolution, corresponding to different frequencies, in units of thermodynamic temperature.

The variance of the cleaned map is optimally minimized at each scale to obtain the coefficients $\beta_{ij}$ or, equivalently, the quadratic quantity \begin{equation}
\chi^{2}_j = \sum_{p} \left\lbrace \widehat{T}_{j}(p)C^{-1}[\widehat{T}_{j}(p)]^t \right \rbrace, \end{equation} where $C^{-1}$ is the inverse of the covariance matrix calculated as the sum of contributions of the CMB and instrumental noise (both, from the map to be cleaned and the templates).

From the previous discussion, it is obvious that the approach to produce an optimal recovery of the CMB would require a certain knowledge of this signal, via its correlations. However, a more robust estimator, without a priori knowledge of the signal to be estimated, may be built by considering only the instrumental noise correlations.

We have checked, however, that the gain in the CMB recovery, by including the information related to the instrumental characteristics is, in practice, very little. Even more, in some situations (as it is the case of the \textit{WMAP} full resolution data, see section \ref{sec:highresWMAP}) the instrumental noise information is limited to the autocorrelation. Therefore, in this work we have decided to perform the internal template fitting with uniform weights for all the pixels at each scale, which implies to minimize the following quantity:

\begin{equation}
E_{j} = \sum_{p} \left[ T_{j}(p) - \sum_{i=1}^{N_{t}} \beta_{ij}t_{ij}(p)\right]^2.
\end{equation}

Finally, we recover a single map performing the wavelet synthesis. It can be written as \begin{equation}
\widehat{T}(\vec{x}) = T(\vec{x}) - \sum_{i=1}^{N_{t}} \sum_{j=1}^{N_{res}} \gamma_{ij}(\vec{x})t_{ij}(\vec{x}), \end{equation} where $N_{res}$ denotes the number of involved resolutions and $\gamma_{ij}$ some new coefficients given as linear combinations of $\beta_{ij}$ coefficients which are the result of the synthesis process.


\section{Analysis of low resolution \textit{WMAP} data}
\label{sec:lowresWMAP}

The instrumental noise in \textit{WMAP} polarization is known to be correlated \citep{Jarosik2011}. Although the \textit{WMAP} data are typically given at a \textit{HEALPix} resolution of $N_{side}=512$, a more accurate version of the pixel-to-pixel correlation is only available at low resolution, namely, $N_{side}=16$. Taking into account this difference, we have performed the cleaning of the \textit{WMAP} data in two cases: for low and high resolution maps. In this section, we analyse the maps at $N_{side}=16$. 

The \textit{WMAP} data are composed by, at least, a superposition of CMB, synchrotron and thermal dust emissions. The \textit{WMAP} team proposed a template fitting in the pixel space to clean the foreground emission in the Ka, Q, V and W maps, using as templates the K band (for the synchrotron) and a low resolution version of the \citet{Finkbeiner1999} model for the thermal dust, with polarization direction derived from starlight measurements \citep{Gold2011}.

In our approach, we use only a synchrotron template, constructed as K-Ka. The reason for neglecting the thermal dust template is  because a previous analysis in real space shows that its coefficients are much smaller than the corresponding ones for the synchrotron template. We clean \textit{Q} and \textit{U} polarization components independently minimizing the variance of the cleaned maps of the Q1, Q2, V1 and V2 differencing assemblies (DAs). The wavelet decomposition is carried out down to resolution $j=3$ for the data map, thus, in addition to the approximation, we have a single detail map at $j=4$. Best fitting coefficients for the considered DAs are given in table \ref{tab:DAcoeff}. We apply the \textit{WMAP} polarization analysis mask that excludes a $26\%$ of the sky.  

\subsection{Cleaned maps}
\begin{figure*}
\centering
\includegraphics[scale=0.37]{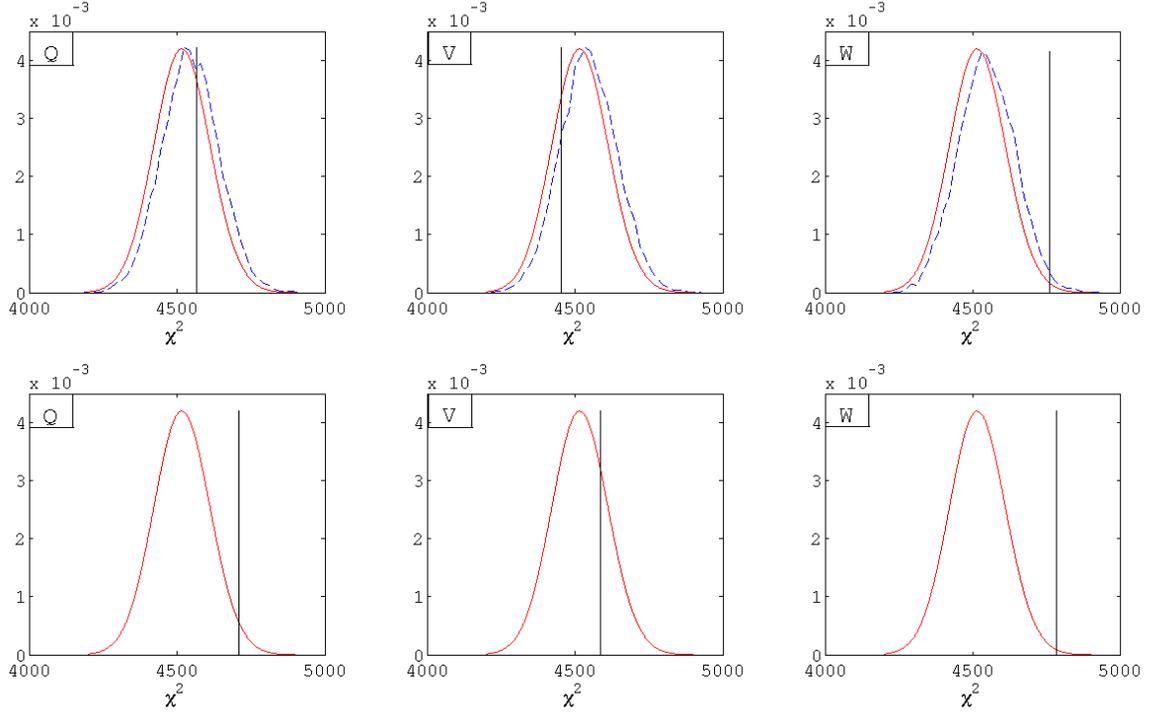} 
\caption{Upper panels show the $\chi^2$ distributions of our cleaned maps and the bottom ones present the same results for the \textit{WMAP} procedure. The solid line (red) corresponds to the theoretical curve of a $\chi^2$ with as many degrees of freedom as pixels outside the mask (i.e., the effective number of pixels in Q and U maps: 4518). The dashed line is the distribution calculated from simulations of our cleaned maps. The vertical line shows the $\chi^2$ value of the data maps in each case. The columns correspond to different frequency bands, from left to right: Q, V and W bands.} 
\label{fig:distributionbands}
\end{figure*}
Since the CMB polarization signal is clearly subdominant in the \textit{WMAP} low resolution data, it is hard to establish a criterion to evaluate the goodness of the cleaning process, and to perform comparisons with different solutions.

We have decided to evaluate this goodness by comparing the cleaned map with the expected signal for a noisy sky following the \textit{WMAP} instrumental noise characteristics. In this sense, a good compatibility with the noise properties would indicate that foregrounds have been satisfactorily reduced.

We generate a set of $10^4$ simulations of the noise maps resulting from our MITC method at Q, V and W frequency bands, $M_r(p)$, with $r \in \{1, ..., 10^4 \}$, in order to construct a $\chi^2$ distribution, calculating each value as \begin{equation} \chi_r^2=\sum_{p,q} M_r(p)N_{Obs}^{-1}(p,q) M_r^t(q),  \end{equation} where $N_{Obs}$ is the noise correlation matrix. A number of simulations of the order of a million is required to estimate this matrix so that the distribution converges to the theorical curve of a $\chi^2$ distribution with as many degrees of freedom as pixels outside the mask in \textit{Q} and \textit{U} maps (in this case, we have 4518 degrees of freedom). This distribution characterizes the expected noise level at each frequency map. We can associate the $\chi^2$ value of the data with relative levels of signal. We can say that the cleaned maps contain more than just noise (typically foreground residuals, since the CMB is subdominant compared to the noise at these scales) if the data value is much higher than typical values of the distribution. Conversely, we can ensure that our maps are compatible with the expected noise and that residuals are small if the data value falls within the distribution. The $\chi^2$ values for each band are listed in table \ref{tab:chisquarebands} and for each DA in table \ref{tab:chisquareDA}.

\begin{table}
\centering
\begin{tabular}{cccc}
\hline
\hline
Frequency band & Q & V & W \\
\hline
$\chi^2_{Cleaned}$ & 4566 & 4453 & 4762  \\ 
$\chi^2_{Forered}$ & 4709 & 4586 & 4787  \\    
\hline
\end{tabular}
\caption{Different values of $\chi^2$ computed with our seven-year cleaned maps per frequency band ($\chi^2_{Cleaned}$) and with \textit{WMAP} seven-year foreground reduced maps per frequency band ($\chi^2_{Forered}$).} \label{tab:chisquarebands} 
\end{table}

\begin{table}
\centering
\begin{tabular}{ccccc}
\hline
\hline
DA & Q1 & Q2 & V1 & V2 \\
\hline
$\chi^2_{Cleaned}$ & 4489 & 4546 & 4405 & 4486  \\ 
\hline
\end{tabular}
\caption{Different values of $\chi^2$ computed with our seven-year cleaned maps per DA.} \label{tab:chisquareDA} 
\end{table}

Our test is based on the assumption that the CMB contribution is negligible. We have tested that the CMB provides a very small contribution (a shift of $\sim 10$ units of $\chi^2$) by generating $10^4$ simulations with CMB and instrumental noise of the cleaned maps. These simulations have been used to compute another $\chi^2$ distribution with the matrix that we have already calculated with only the noise component. When distributions are compared with each other we observe this typical deviation. Thus, the CMB contribution to the value of the $\chi^2$ of the data is negligible and, therefore, any significant deviation from the mean value has to be assigned to foreground residuals.

\begin{figure}
\centering
\includegraphics[scale=0.45]{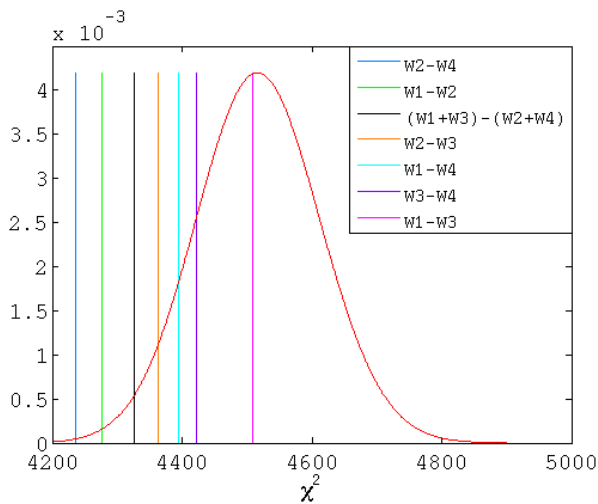} 
\caption{The combinations of the raw W-band maps. Theoretical curve of the $\chi^2$ distribution is represented by a solid red line and the $\chi^2$ values are shown by successive vertical lines, from left to right: W2-W4, W1-W2, (W1+W3)-(W2+W4), W2-W3, W1-W4, W3-W4 and W1-W3.} 
\label{fig:Wnoise}
\end{figure}

An indirect comparison can be made between the \textit{WMAP} procedure and our MITC method through the relative possitions of the $\chi^2$ value of the data with respect to the distribution. As seen in figure \ref{fig:distributionbands}, we obtain that the value of the cleaned maps is fully compatible with instrumental noise at Q and V frequency bands. At W band the $\chi^2$ value of the cleaned data is in the tail of the distribution probably due to the presence of foreground residuals. The deviation is even larger when the \textit{WMAP} procedure is used. A significant improvement is also found at Q band since the $\chi^2$ value is shifted from $2 \sigma$ to $0.5 \sigma$ when our MITC method is used. 

In addition, although we use a template that is noisier than the ones used by the \textit{WMAP} team, the noise levels of our cleaned maps are lower. We have measured a difference of about a $10\%$ in terms of the standard deviation of the data maps (this difference is confirmed by instrumental noise simulations).

In order to check further the apparent excess of signal at W band obtained by the two approaches, we computed analytically the noise covariance matrix of different combinations of the raw W-band DAs maps which contain, in principle, only a combination of instrumental noise. With these covariance matrices, based on the full-sky covariance matrices of each DA, a $\chi^2$ value of the data maps is obtained. It is shown in figure \ref{fig:Wnoise} that these maps are still compatible with the expected noise. However, it is significant that all values are to the left of the distribution and that the most deviated ones involve W2, followed by W4. We have also analysed the distribution of the $\chi^2$ values of the single-year foreground reduced maps supplied by the \textit{WMAP} team for each DA at the W band, obtaining values more deviated towards the tails for the W2 and W4 DAs. This may suggest a not good enough characterization of the instrumental noise for these DAs.

\subsection{Polarization power spectra, $C_{\ell}^{EE}$ and $C_{\ell}^{BB}$}
We carry out an estimation of the polarization spectrum using our cleaned maps of the Q1, Q2, V1 and V2 DAs. A pseudo cross-power spectrum $\hat{D}_{\ell}^{AB}$ between any two differencing assemblies A and B can be calculated as \begin{equation}
\hat{D}_{\ell}^{AB} = \sum_{\ell'} M_{\ell \ell'}^{AB} |p_{\ell'}|^{2}B_{\ell'}^A B_{\ell'}^B \langle C_{\ell'}^{AB} \rangle + \langle N_{\ell}^{AB} \rangle ,
\label{eq:pseudocls}\end{equation} where $A, B \in \{Q1, Q2, V1, V2 \ | \ A \neq B\}$; and, in the case of an EE power spectra, \begin{equation}
\hat{C}_{\ell}^{AB} = \dfrac{1}{2\ell + 1}\sum_{m=-\ell}^{\ell} e_{\ell m}^A e_{\ell m}^{B*},
\end{equation} where $e_{\ell m}$ are the spherical harmonic coefficients of the E-mode. Assuming a circular beam response, we denote the beam of the A map as $B_{\ell}^A$ and the window function of the \textit{HEALPix} pixel by $p_{\ell}$; $\langle N_{\ell}^{AB} \rangle$ is the noise cross-power spectrum. The bias introduced by this term comes from the internal template fitting procedure. It is small and controled by simulations. Finally, the coupling kernel matrix $M_{\ell \ell'}$ is described in \citet{Hivon2002} and, for the case of the polarization components, in Appendix A of \citet{Kogut2003}. This procedure is usually referred to as MASTER estimation.
An estimator, $\hat{C}_{\ell}$, can be computed as a linear combination of the six different spectra weighted by the inverse of their variances in the following way:
\begin{equation}
\hat{C}_{\ell} = \left(\sum_i \dfrac{1}{\sigma^2_i}\right)^{-1}\sum_i \dfrac{1}{\sigma^2_i}\hat{C}_{\ell}^i,
\label{eq:comb}\end{equation} where $i = AB$ and $\sigma^2_i = \sigma_A \sigma_B$. These variances are given by the \textit{WMAP} team in the \textit{LAMBDA} web site\footnote[1]{http://lambda.gsfc.nasa.gov/}. 

\begin{figure}
\centering
\includegraphics[scale=0.40]{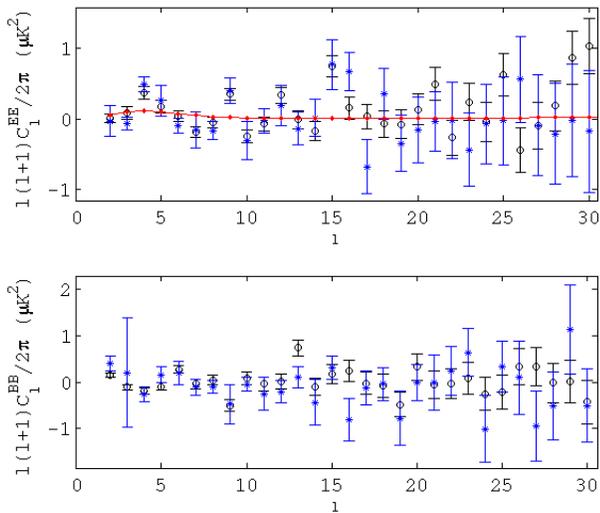} 
\caption{Polarization power spectrum EE (upper pannel) and BB (bottom pannel) for low resolution analysis. Circles are the spectrum supplied by \textit{WMAP} team and asterisks represent our estimation. The fiducial model is plotted by the solid line. } 
\label{fig:Spelowres}
\end{figure}

The resulting power spectra are shown in figure \ref{fig:Spelowres}. From the $C_{\ell}^{EE}$ spectrum we can say that most of the values are compatible with zero, so there is almost no signal except perhaps for low multipoles $\ell \lesssim 6$. As expected, the B-mode spectrum $C_{\ell}^{BB}$ signal is compatible with zero. Both spectra are compatible with those that the \textit{WMAP} team supplies. Our error bars are larger than those obtained by the \textit{WMAP} team, because of the use of an estimator that is not optimal, a pseudo-spectrum, whereas the \textit{WMAP} team uses a pixel-base likelihood. 


\section{Analysis of high resolution \textit{WMAP} data}
\label{sec:highresWMAP}

\begin{figure}
\centering
\includegraphics[scale=0.37]{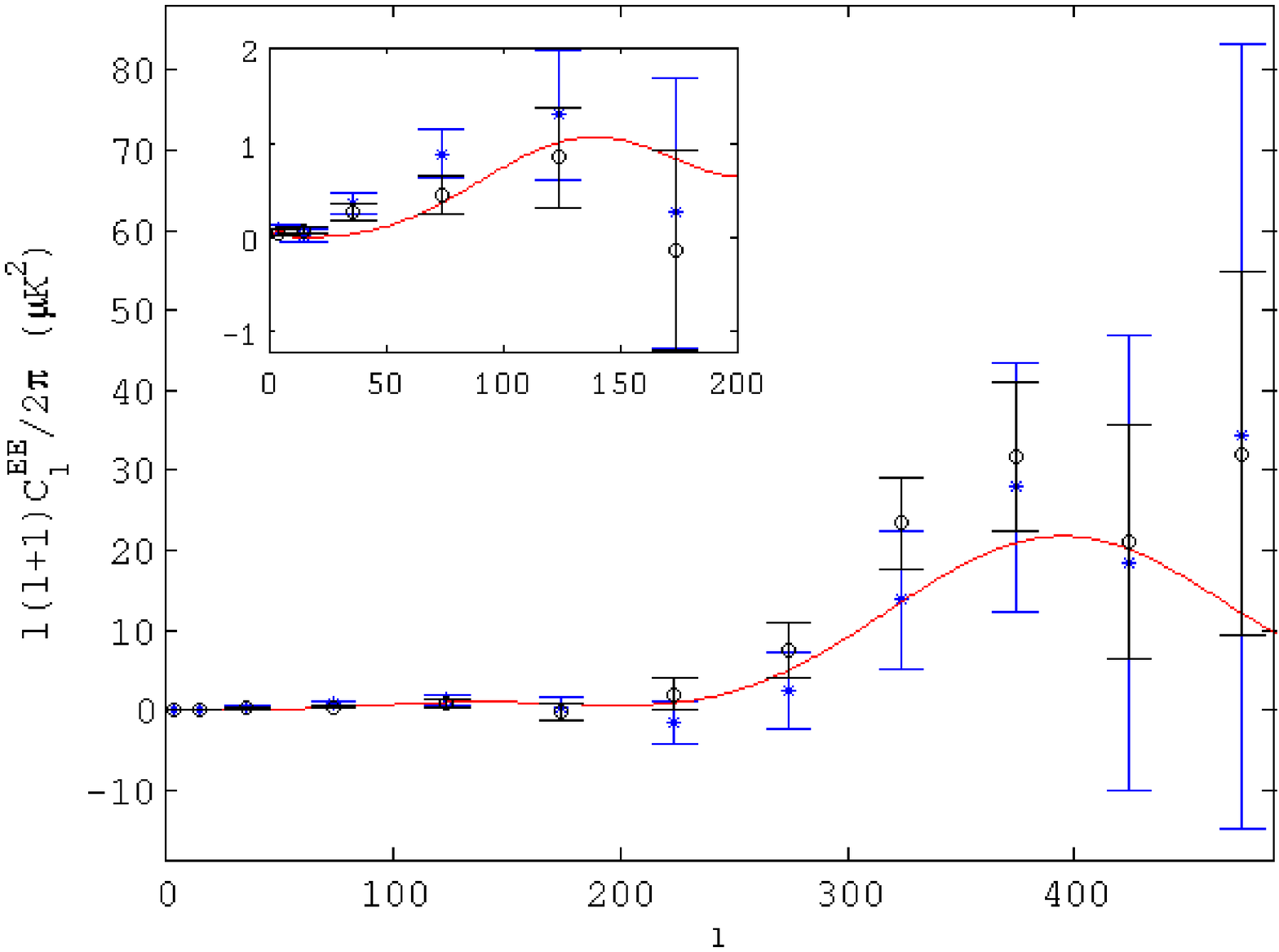} 
\includegraphics[scale=0.37]{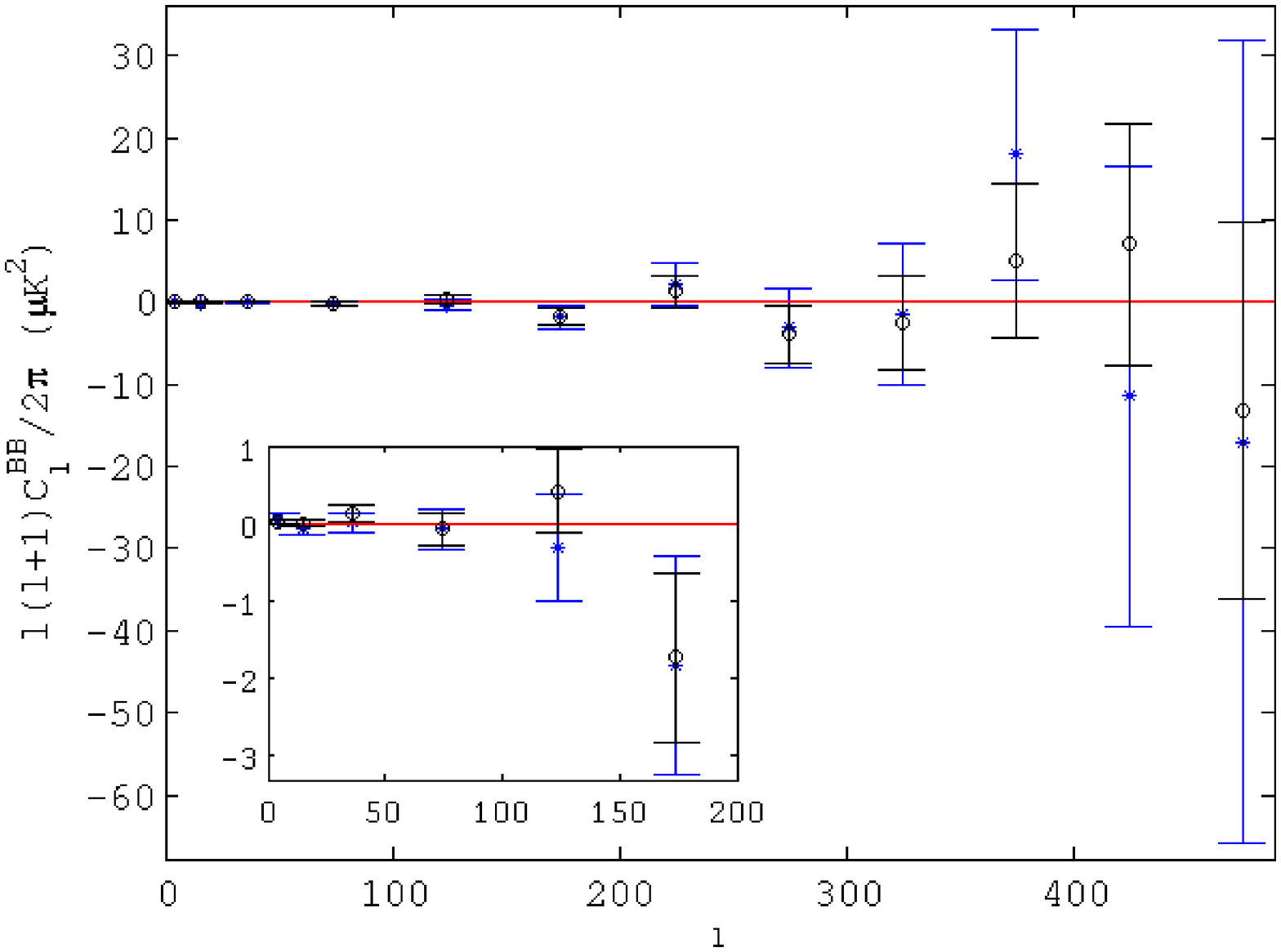} 
\caption{Polarization power spectrum EE (upper pannel) and BB (bottom pannel) for high resolution analysis. The circles are the spectrum supplied by the \textit{WMAP} team and the asterisks represent our estimation. The solid line represents the fiducial model for $C_{\ell}^{EE}$ and the zero value for $C_{\ell}^{BB}$. } 
\label{fig:Spehighres}
\end{figure}

In this section, we analyze \textit{WMAP} data maps at $N_{side}=512$. This approach allows us to study the cleaning at smaller scales where, \textit{a priori}, the correlation of the noise is less important. So then we only take into account the noise covariance matrix of each pixel. In this case, the cleaning method based on the wavelet space is applied using two different internal templates. The first one is constructed as K-Ka and accounts for the synchrotron radiation. The second one is built as V1-W3 and attempts to characterize the thermal dust. The V and W DAs to be cleaned have been selected by having a lower noise. As in the low resolution case, the wavelet decomposition is carried out until resolution $j=3$ for the approximation map, hence we have additionally 6 different detail maps in this high resolution case. Again, the \textit{WMAP} polarization mask is used. Best fitting coefficients are shown in table \ref{tab:DAcoeffhigh}.

From the cleaned \textit{Q} and \textit{U} maps we study the power spectra, $C_{\ell}^{EE}$ and $C_{\ell}^{BB}$, as in the previous section. The spectrum error bars are estimated from $10^3$ noise simulations. In general, the CMB contribution is neglegible compared to the noise one. Bins are taken as a weighted average of the multipoles involved. These weights are calculated as the inverse of the variance of each $C_{\ell}$.

\begin{table*}
\centering
\begin{tabular}{ccccccccc}
\hline
\hline
DA & Template & Stokes & Q1 & Q2 & V2 & W1 & W2 & W4 \\
\hline
\multirow{4}{*}{Detail ($j=9$)} & {\multirow{2}{*}{(K-Ka)}} & \textit{Q} Stokes & -0.0175 & -0.0067 & 0.0024 & 0.0754 & 0.0072 & -0.0222 \\ 
                             & & \textit{U} Stokes & 0.0525 & 0.0173 & 0.0085 & -0.0869 & 0.1171 & -0.0562 \\
 & {\multirow{2}{*}{(V1-W3)}} & \textit{Q} Stokes & -0.0008 & 0.0023 & -0.0015 & 0.0013 & -0.0015 & 0.0054 \\ 
                                         & & \textit{U} Stokes & 0.0007 & 0.0006 & 0.0022 & 0.0005 & -0.0226 & -0.0106 \\
\hline
\multirow{4}{*}{Detail ($j=8$)} & {\multirow{2}{*}{(K-Ka)}} & \textit{Q} Stokes & -0.0238 & 0.0185 & 0.0135 & -0.0046 & 0.0076 & 0.0311 \\ 
                             & & \textit{U} Stokes & -0.0080 & -0.0176 & -0.0059 & -0.0513 & 0.0138 & -0.0196 \\
 & {\multirow{2}{*}{(V1-W3)}} & \textit{Q} Stokes & -0.0044 & -0.0003 & -0.0008 & -0.0048 & 0.0020 & 0.0071 \\ 
                                         & & \textit{U} Stokes & 0.0012 & 0.0028 & 0.0042 & 0.0013 & 0.0030 & 0.0012 \\
\hline
\multirow{4}{*}{Detail ($j=7$)} & {\multirow{2}{*}{(K-Ka)}} & \textit{Q} Stokes & 0.0006 & 0.0045 & -0.0091 & -0.0227 & 0.0288 & -0.0224 \\ 
                             & & \textit{U} Stokes & -0.0208 & 0.0196 & 0.0038 & -0.0263 & 0.0047 & -0.0142 \\
 & {\multirow{2}{*}{(V1-W3)}} & \textit{Q} Stokes & 0.0054 & 0.0004 & -0.0002 & -0.0053 & -0.0002 & 0.0069 \\ 
                                         & & \textit{U} Stokes & -0.0017 & 0.0006 & 0.0032 & 0.0019 & -0.0036 & 0.0009 \\
\hline
\multirow{4}{*}{Detail ($j=6$)} & {\multirow{2}{*}{(K-Ka)}} & \textit{Q} Stokes & 0.0224 & -0.0112 & -0.0073 & 0.0302 & -0.0057 & 0.0407 \\ 
                             & & \textit{U} Stokes & 0.0278 & 0.0172 & 0.0104 & 0.0144 & 0.01797 & 0.0076 \\
 & {\multirow{2}{*}{(V1-W3)}} & \textit{Q} Stokes & 0.0079 & 0.0014 & -0.0039 & -0.0070 & -0.0095 & -0.0038 \\ 
                                         & & \textit{U} Stokes & 0.0066 & -0.0051 & 0.0016 & -0.0012 & -0.0096 & 0.0179 \\
\hline
\multirow{4}{*}{Detail ($j=5$)} & {\multirow{2}{*}{(K-Ka)}} & \textit{Q} Stokes & 0.0351 & 0.0729 & 0.0702 & -0.0325 & 0.0253 & -0.1171 \\ 
                             & & \textit{U} Stokes & 0.0567 & 0.0523 & 0.0670 & 0.1166 & 0.0255 & -0.0225 \\
 & {\multirow{2}{*}{(V1-W3)}} & \textit{Q} Stokes & -0.0131 & 0.0012 & 0.0207 & -0.0070 & -0.0578 & 0.0508 \\ 
                                         & & \textit{U} Stokes & -0.0080 & 0.0009 & -0.0071 & -0.0362 & -0.0006 & 0.0363 \\
\hline
\multirow{4}{*}{Detail ($j=4$)} & {\multirow{2}{*}{(K-Ka)}} & \textit{Q} Stokes & 0.2292 & 0.0571 & 0.0548 & -0.0835 & 0.3029 & -0.2239 \\ 
                             & & \textit{U} Stokes & 0.0741 & 0.1172 & -0.0844 & 0.1445 & -0.2241 & -0.2723 \\
 & {\multirow{2}{*}{(V1-W3)}} & \textit{Q} Stokes & -0.0708 & 0.0774 & -0.1389 & 0.0386 & 0.1316 & -0.0135 \\ 
                                         & & \textit{U} Stokes & -0.0440 & 0.0359 & -0.0397 & 0.1985 & -0.2028 & 0.2320 \\
\hline
\multirow{4}{*}{Approximation ($j=3$)} & {\multirow{2}{*}{(K-Ka)}} & \textit{Q} Stokes & 0.1732 & 0.2976 & 0.1988 & 0.1413 & 0.1234 & 0.2612 \\ 
                             & & \textit{U} Stokes & 0.1397 & 0.2315 & 0.1674 & 0.1607 & 0.4529 & 0.5386 \\
 & {\multirow{2}{*}{(V1-W3)}} & \textit{Q} Stokes & -0.2852 & 0.2030 & 0.6050 & 0.5513 & 0.2601 & -1.1963 \\ 
                                         & & \textit{U} Stokes & 0.2087 & 0.1022 & -0.0048 & 0.5052 & 0.0160 & -0.3867 \\
\hline
\end{tabular}
\caption{Template cleaning coefficients for \textit{Q} and \textit{U} Stokes parameters and different frecuency bands for the high resolution case.} \label{tab:DAcoeffhigh} 
\end{table*}

The spectra are compatible with those that the \textit{WMAP} team has
obtained. Our error bars are larger at high multipoles, since, at this
scale range, the number of effective cross-spectra is much smaller
than the one used by the \textit{WMAP} team (where all the W-band DAs
are available). Nevertheless, as seen in figure \ref{fig:distributionbands}, our cleaned maps seem to present a lower level of contamination than those supplied by the \textit{WMAP} team. We
  expected a better foreground removal since the templates used are
  closer to the foreground signal distribution accross the sky in our
  case. However, whether this may have an impact on the determination of the cosmological parameters is not clear and would require an exhaustive analysis which is outside the scope of this paper. 

Finally, the two points accounting for the largest scales of the spectra are taken from figure \ref{fig:Spelowres}, since the correlation of the noise at these scales is important, and it has been better modeled in the previous section, where a more accurate version of this information was available.

The resulting power spectra are presented in figure \ref{fig:Spehighres}. Our outcome is compatible with the \textit{WMAP} team analysis, where it is possible to distinguish the acoustic peak around $\ell \sim 400$ in the E-mode spectrum. As expected, the B-mode power spectrum is compatible with zero. 

Our independent approach can be seen as a confirmation of the previously detection reported by the \textit{WMAP} team.

\begin{figure}
\centering
\includegraphics[scale=0.25]{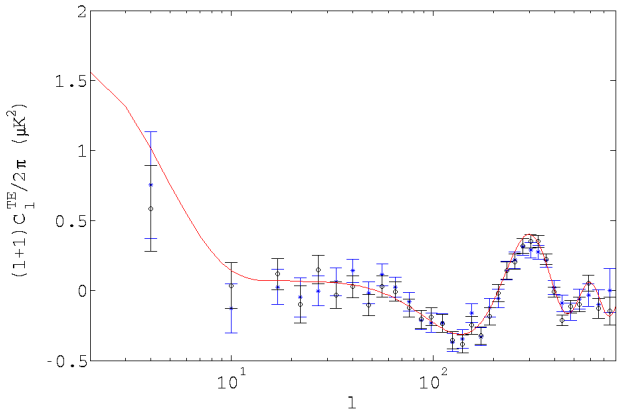} 
\caption{Polarization power spectrum TE. Circles are the spectrum supplied by \textit{WMAP} team and asterisks represent our estimation. The fiducial model is plotted by the solid line. } 
\label{fig:SpeTE}
\end{figure}

A similar procedure is applied to determine the correlation between temperature and E-mode polarization data. We carry out the analysis using our cleaned maps of the \textit{Q} and \textit{U} Stokes parameters and the foreground reduced temperature maps that the \textit{WMAP} team supplies. In this case, the combination of two maps of the same DA is allowed and the equation \ref{eq:pseudocls} has the same form as long as we add that, in $C_{\ell}^{AB}$, $A$ refers to the DA of temperature maps and $B$ to the DA of polarization maps, with $A, B \in \{Q1, Q2, V2, W1, W2, W4 \}$. For the temperature maps, we use the temperature analysis mask that the \textit{WMAP} team supplies and the MASTER estimation is computed as is described in \citet{Kogut2003}.

Since the CMB cosmic variance in temperature contributes significantly, we cannot ignore it this time in the calculation of the error bars. So we generate $10^3$ simulations of CMB plus noise, which undergo the same process of cleaning and combination to get the cross power spectrum, with which we estimate the error bar. These errors are relatively larger with respect to the \textit{WMAP} results, due to the less number of effective cross spectra. In addition, we remark that, while we use a pseudo-spectrum, low resolution analysis is performed through a pixel-base likelihood by the \textit{WMAP} team.

The resulting cross power spectrum is presented in figure \ref{fig:SpeTE}, and it is compatible with the power spectrum that the \textit{WMAP} team supplies.

\section{Conclusions}
\label{sec:conclusions} 
We introduce an internal template cleaning method that uses a wavelet decomposition on the sphere. Among its advantages, it is included the possibility of multi-resolution analysis, allowing an effective variation of the spectral index in the sky. Much lower computational time is needed than with other widely-used continous wavelets. In addition, a good treatment of incomplete sky coverage is given because of the compact support of the \textit{HEALPix} wavelet.

The MITC method result is a set of some cleaned maps at several frequencies that can be used, for instance, to verify whether any detected feature of the data is actually monochromatic or
not. The exclusive use of internal templates allows us to analyze the maps without making any prior assumptions about the foregrounds in polarization. However, although the implementation that is shown in this work make use of only internal templates, it can be trivially extended to deal with external templates as well.

We perform an analysis of 7-year \textit{WMAP} data obtaining outcomes that are compatible with the \textit{WMAP} team results. Furthermore, we have hints of better cleaning of the Q-band map at large scales and we have obtained cleaned maps, at least, as good as those that the \textit{WMAP} team supplies for V and W bands. Let us remark that our approach does not make use of any additional template: everything is obtained from \textit{WMAP} data. We have checked that, although we use noisier templates, the instrumental noise levels of the final cleaned maps are lower than those of the maps provided by the \textit{WMAP} team.

High resolution maps are also analysed. In agreement with the \textit{WMAP} team, we find an E-mode detection at $\ell \sim 400$, as predicted by the standard $\Lambda CDM$ model. We also obtain that the B-mode level is compatible with zero. These independent findings are a confirmation of the result already presented by the \textit{WMAP} team.

The clean maps produced at this work, both at low and high resolution, are available at the following website: http://max.ifca.unican.es/cobos/WMAP7yrPOL


\section*{acknowledgments}
Authors acknowledge partial financial support from the Spanish \textit{Ministerio de Ciencia e Innovaci\'on} Projects AYA2010-21766-C03-01 and Consolider-Ingenio 2010 CSD2010-00064. RFC thanks financial support from Spanish CSIC for a \textit{JAE-predoc} fellowship. PV thanks financial support from the Ram\'on y Cajal program. The authors acknowledge the computer resources, technical expertise and assistance provided by the \textit{Spanish Supercomputing Network} (RES) node at Universidad de Cantabria. We acknowledge the use of \textit{Legacy Archive for Microwave Background Data Analysis} (LAMBDA) and the assistance provided by Benjamin Gold by e-mail. The HEALPix package was used throughout the data analysis \citep{Gorski2005}.
\bibliographystyle{mn2e}
\bibliography{cites}

\end{document}